# Cascaded frequency conversion of highly charged femtosecond spatiotemporal optical vortices

Qingqing Liang,[1] Qiyuan Zhang,[1] Dan Wang,[1] Enliang Zhang,[1] Jianhua Hu,[1] Ming Wang,[1] Jinxin Wu,[1] Grover A. Swartzlander,[2] Yi Liu,[1, *]

[1]*Shanghai Key Lab of Modern Optical System, University of Shanghai for Science and Technology, 516, Jungong Road, 200093 Shanghai, China*
[2]*Chester F. Carlson Center for Imaging Science, Rochester Institute of Technology, Rochester, New York, USA*

*[*] yi.liu@usst.edu.cn*

The degrees of freedom inherent in spatiotemporal optical vortices (STOV's) afford intriguing opportunities to manipulate complex light fields for broad applications such as optical communication, light-matter interactions, particle manipulation, quantum optics, and electron acceleration in the relativistic regime. Unlike previous studies examining the second harmonic generation (SHG) of STOV's having an input topological charge (TC) $l^{(\omega)} = 1$, here we experimentally demonstrate cascaded second and third harmonic generation of STOV's to achieve more diverse wavelength selectivity via sum frequency generation (SFG), achieving unprecedented TC values up to $l^{(\omega)} = 40$. The large TC values are attributed to second (third) harmonic generation of an incident beam satisfying $l^{(2\omega)} = 2l^{(\omega)}$ ($l^{(3\omega)} = 3l^{(\omega)}$). What is more, the wavelength of the generated STOV field was found to be tunable by controlling the position of the phase singularity in the frequency domain, and optimizing the nonlinear phase matching condition. Our experimental measurements extend the principle of conservation of the spatiotemporal topological charge to general nonlinear optical parametric processes, suggesting a fundamental approach to the production of STOV fields of arbitrarily large TC and at arbitrary visible wavelengths, and beyond (e.g., the ultraviolet range).

Angular momentum degrees of freedom afford opportunties to shape complex light fields, and by extension, allow intriguing ways to manipulation the interaction of light with matter [1-4]. Whereas the spin component of angular momentum is associated with atomic transitions and polarization elements such as wave retarders, the orbital angular momentum is associated with the circulation of the optical phase gradient (i.e., vortices) $\oint \vec{\nabla}\phi \cdot \vec{dr} = 2\pi l$ where $\phi$ is the value of optical phase along the closed integration path and $l$ is an integer called the topological (or vortex) charge. Since first proposed in the context of coherent laser beams in 1989 [1], the study of OAM phenomena has bloomed and its applications have spanned diverse areas such as high capacity optical communication [5, 6], optical trapping[7-9], superresolution microscopy[10-12], quantum key distribution[13], and exoplanet detection [14].

Recently spectral filtering techniques have been used to imprint vortex phase into the temporal domain of a coherent laser beam, resulting in spatiotemporal optical vortices (STOV's) [15, 16]. In this case the spiral phase manifests in the space and time plane, and the electric field may be described by:

$$E(x,y,z,\tau) = a(\tau/\tau_s + i\,\text{sgn}(l)\,x/x_s)^{|l|}E_0(x,y,z,\tau)$$

$$= A(x,\tau)e^{il\Phi_{st}}E_0(x,y,z,\tau), \qquad (1)$$

where $\tau = t - z/v_g$ is a time coordinate moving at the speed of light, $v_g$ is the group velocity, $\tau_s$ and $x_s$ are temporal and spatial widths of the STOV field, $\Phi_{st} = l\tan^{-1}(x\tau_s/tx_s)$ is the space-time phase whose gradient in the $x - \tau$ plane circulates, $a = \sqrt{2}[(x_0/x_s)^2 + (\tau_0/\tau_s)^2]^{-1/2}$ is a normalization factor, $A(x,\tau) = a[(x/x_s)^2 + (\tau/\tau_s)^2]^{|l|/2}$, and $E_0$ describes a spatio-temporal Gaussian pulse characterized by 1/e width parameters $x_0$ and $\tau_0$ [16].

This newly demonstrated degree of freedom for the optical field has attracted considerable attention owing to the new opportunities for optical field manipulation [17, 18], light-matter interaction[19], as well as information transfer [20]. For example, it has been theoretically demonstrated that the accelerated electrons can be trapped in the spatiotemporal singularity during the interaction of plasma with intense STOV fields ($I \geq 10^{18}\,\text{W}/\text{cm}^2$) in the relativistic regime, producing attosecond electron sheet with potential applications in attosecond electron diffraction and isolated attosecond pulse generation [19].

Shortly after the demonstration of the STOV field in the near-infrared regime [21, 22], its frequency conversion has been studied to extend its wavelength range. G. Gui *et al.* employed a beta barium borate (BBO) crystal to produce the second harmonic of the STOV field and observed an efficient conversion [21]. In their study, the TC of the incident STOV field was limited to $l^{(\omega)} = 1$ and the generated second harmonic processes a spatiotemporal TC of $l^{(2\omega)} = 2$, with the topological charge conserved (i.e., two photons having $l^{(\omega)} = 1$ at the fundamental wavelength were converted to a single photon having $l^{(2\omega)} = 2$ at the second harmonic wavelength). The frequency doubling of STOV fields having $l^{(\omega)} > 2$ were unexplored. Moreover, the transfer of topological charge in other optical parametric processes such as sum frequency generation, which can extend the wavelength of the STOV field into the UV and deep UV range, was not considered and rules for the transfer of TC remained unknown.

In this study, we experimentally demonstrated the cascaded second harmonic generation and sum frequency generation (SFG) of femtosecond STOV fields having very large topological charges, e.g., $l^{(\omega)} > 40$. With BBO crystals of different thickness, we first obtained frequency doubling of STOV field for various values of TC. It was found that the central wavelength of the second harmonic STOV field can be tuned by adjusting the spectral position of the vortex singularity of the fundamental optical field and optimized via phase matching of the SHG process. With another SFG BBO crystal cascaded after the SHG crystal, the third harmonic of the fundamental STOV field was achieved, with its TC tripled with respect to the incident light field. Degradation of topological charge of the third harmonic was observed for thick SFG crystal, which indicates the crucial role of the group delay dispersion between the FW and the SH pulses during the SFG process. Our findings demonstrate that the OAM of the STOV field can be well conserved during the SHG and SFG processes even for very high-order topological charges, and provides robust methods for the generation of femtosecond STOV fields for wavelengths across the UV-visible spectrum.

**Results**

**Fundamental STOV generation and characterization**

We generated the fundamental 800 nm femtosecond STOV field using a 4f pulse shaper integrated with a spatial light modulator (SLM) [15], as schematically presented in Fig. 1. The resulting STOV field was characterized at the pulse shaper exit via the diffraction method established by S. L. Huang *et al*. [23] (experimental details are provided in the Methods section). Figure 2 displays diffraction patterns for the fundamental STOV field with topological charges reaching up to $l^{(\omega)} = 14$. Unlike conventional beams lacking orbital angular momentum ($l^{(\omega)} = 0$), the STOV diffraction patterns exhibit multiple lobes intersected by tilted dark stripes. These stripes arise from interference between optical field components carrying opposite phases, with the number of dark fringes matching the STOV's topological charge [23]. Supplementary Fig. 1 presents calculated diffraction patterns for varying topological charges, demonstrating quantitative agreement with experimental observations and confirming successful generation of the fundamental 800 nm STOV field.

We further examined the spectral properties of the STOV field by scanning a fiber tip transversely (*y*-direction) across the 800 nm beam. Results for $l^{(\omega)} = 4$ are shown in Fig. 3a, revealing a distinct spatial-spectral chirp characteristic of STOV pulses. In contrast, measurements for $l^{(\omega)} = 0$ (Supplementary Fig. 2a) show spectra centered uniformly at 800 nm without spatial chirp.

**Second-harmonic generation (SHG) of STOVs**

When the fundamental STOV field at 800 nm illuminates the BBO crystal, it generates a second harmonic field at 400 nm. Figure 4 presents the diffraction patterns for this second harmonic field. For fundamental pulses carrying spatiotemporal topological charges, we find that the second harmonic exhibits corresponding topological charges with values doubled (i.e., $l^{(2\omega)} = 2l^{(\omega)}$). For example, an 800 nm pulse with $l^{(\omega)}$ =10 produces a second harmonic with $l^{(2\omega)}$=20. We examined topological charges up to $l^{(\omega)}$=30 and consistently observed clear evidence of topological charge doubling. Figure 4b shows a magnified diffraction pattern for $l^{(2\omega)}$=60. While further increases in topological charge were limited by the spatial resolution of our CCD camera, we successfully implemented fundamental pulses with topological charges up to $l^{(\omega)}$=40.

We further characterized the spectrum of the second harmonic STOV. As shown in Fig. 3b, spectral shifts along the *y*-direction reflect the characteristic spatial and spectral properties of STOV fields [15,16]. Using BBO crystals of different thicknesses for SHG (Supplementary Fig. 3), we observed that thicker crystals (e.g., 1 mm) cause diffraction pattern blurring. This observation agrees with previous reports of spatiotemporal distortion in 400 nm STOV fields with thicker BBO crystals, attributed to spatiotemporal astigmatism between the 800 nm and 400 nm components within the nonlinear crystal [21].

Figure 5a presents the energy dependence of the second harmonic STOV field. Experimental data follow a quadratic scaling law ($I_{SHG} \sim \chi^{(2)}_{sum} I^2_{800}$), consistent with second-order nonlinear optical processes. For larger $l^{(\omega)}$ values, we observed gradually decreasing conversion efficiency (Fig. 5a), attributable to temporal lengthening of the fundamental pulse and increased transverse beam size at higher topological charges [24], which collectively reduce the 800 nm laser intensity $I_{800}$.

Leveraging our femtosecond laser system's broad spectral bandwidth (30 nm FWHM), we demonstrate tunability of the second harmonic STOV's central wavelength by adjusting the spectral location of the phase singularity in the fundamental STOV while optimizing the BBO crystal's azimuthal angle. Supplementary Fig. 5 shows diffraction patterns of 800 nm STOV fields with phase singularities at different spectral positions, alongside corresponding second harmonic diffraction patterns and spectra. When we positioned the phase singularity at different locations on the SLM, the fundamental STOV's diffraction pattern became asymmetric, with its central dark zone shifting between longer and shorter wavelengths. After optimizing phase matching in a thin BBO crystal centered at the singularity wavelength, we consistently generated high-quality second harmonic STOVs (middle row, Supplementary Fig. 5). The second harmonic spectra systematically shift from 393 nm to 409.5 nm, demonstrating spectral flexibility through simple SLM and crystal adjustments.

**Third-harmonic generation (THG) via sum-frequency mixing**

We next generated the third harmonic via sum frequency generation (SFG) using a type I BBO crystal with the fundamental and second harmonic STOV fields. Crystal thickness proved critical, as group velocity dispersion induces temporal separation between the 800 nm and 400 nm fields in thicker crystals. We selected $d$ = 0.1 mm crystals to minimize temporal walk-off. For SFG with type I crystal, parallel polarization of both input fields is ideal. However, our type I SHG BBO produces orthogonally polarized fundamental and second harmonic fields for optimal SHG efficiency. We therefore adjusted the SHG crystal's azimuthal angle to generate parallel polarization components in the 800 nm and 400 nm outputs [25], accepting reduced SHG conversion efficiency. Figure 6 presents diffraction patterns of the third harmonic STOV. For an incident 800 nm STOV with topological charge $l^{(\omega)}$=4, we observed $l^{(3\omega)}$=12 in the third harmonic, confirming the relationship $l^{(3\omega)} = l^{(\omega)} + l^{(2\omega)} = 3l^{(\omega)}$. These results establish topological charge conservation in SFG across all tested charges.

Spectral characterization of the third harmonic STOV (Fig. 3c) revealed characteristic spatial chirp along the *y*-direction for $l^{(3\omega)}$=12. Control measurements without spatial chirp ($l^{(\omega)}$=0) show uniform spectra (Fig. 3d). Figure 5b displays cubic energy scaling ($I_{THG} \sim \chi^{(3)}_{sum} I^3_{800}$) for $l^{(3\omega)}$=3,6, and 9, consistent with third-order nonlinear processes.

**Discussion**

The doubling of the spatiotemporal topological charge (TC) of STOV fields during second-harmonic generation (SHG) in thin BBO crystals ($l^{(2\omega)} = 2l^{(\omega)}$) has been experimentally observed and thoroughly explained in prior studies [21, 22]. To interpret our observations of TC tripling in cascaded sum-frequency generation (SFG) ($l^{(3\omega)} = 3l^{(\omega)}$; Fig. 6), we examine this process within the framework of coupled-wave equations [26]. Under the paraxial approximation with perfect phase matching, and neglecting pump depletion and group-velocity dispersion, the sum-frequency field scales proportionally to the product of the fundamental and second-harmonic fields:

$$\begin{aligned}
E^{(3\omega)}(x,y,\zeta) &= E_0 e^{il^{(3\omega)}\Phi(x,y,\zeta)} \\
&\propto \chi_{\text{sum}}^{(2)}[E^{(\omega)}(x,y,\zeta)][E^{(2\omega)}(x,y,\zeta)] \\
&\propto \chi_{\text{sum}}^{(2)}[E^{(\omega)}(x,y,\zeta)]\chi_{\text{SHG}}^{(2)}[E^{(\omega)}(x,y,\zeta)]^2 \\
&\propto e^{il^{(\omega)}} e^{il^{(2\omega)}}[E^{(\omega)}(x,y,\zeta)]^3 \\
&\propto e^{i3l^{(\omega)}}[E^{(\omega)}(x,y,\zeta)]^3
\end{aligned} \quad (2)$$

Consequently, the topological charge for this cascaded SFG process follows $l^{(3\omega)} = l^{(\omega)} + l^{(2\omega)} = 3l^{(\omega)}$, which accounts for the observed tripling. Equation (2) further corroborates the cubic intensity dependence of the third-harmonic energy shown in Fig. 5b.

For thicker SHG crystals, previous work has demonstrated that the topological structure of the second-harmonic pulse can be altered by complex spatiotemporal astigmatism [21]. In such cases, group-velocity mismatch, pump depletion, as well as absorption and losses must be considered [21]. In our experiments employing thicker BBO crystals (~1 mm) for third-harmonic generation (Supplementary Fig. 4), diffraction patterns became blurred at high topological charges ($l^{(3\omega)} \geq 15$), indicating degradation of the topological charge mode. This effect arises because group-velocity dispersion between the 800 nm and 400 nm components produces a 299 fs temporal delay after propagation through the 1-mm-thick crystal. Since the topological charges of the incident fields circulate within the spatiotemporal domain, partial temporal walk-off between the fundamental and second-harmonic STOV fields results in non-pure topological charges for the generated third harmonic. The reduced diffraction contrast observed for such third-harmonic pulses with mixed TC (Supplementary Fig. 4) occurs because beams carrying different topological charges exhibit distinct numbers of nodes in their diffraction patterns.

Given the conservation of topological charge in SHG and SFG processes, we anticipate that spatiotemporal orbital angular momentum (ST-OAM) will likewise be conserved in other nonlinear processes including optical rectification, direct third-harmonic generation, four-wave mixing, and even high-harmonic generation [27, 28]. These nonlinear optical techniques hold significant potential for extending the accessible wavelength range of STOV fields from the current near-infrared regime into the terahertz, ultraviolet, and extreme-UV spectral regions. Such expansion would create new pathways for diverse photonic applications of STOV fields.

In conclusion, we have demonstrated that cascaded SHG and SFG of 800 nm femtosecond STOV pulses can be achieved using two nonlinear crystals, accommodating incident light fields with arbitrarily large ST-OAM. We have observed general conservation laws governing ST-OAM transfer during SHG ($l^{(2\omega)} = 2l^{(\omega)}$) and SFG ($l^{(3\omega)} = 3l^{(\omega)}$). The energy scaling of the harmonic STOV fields mirrors conventional frequency-doubling and -tripling processes, though reduced conversion efficiency occurs at higher TC values due to diminished light intensity in the STOV field. Furthermore, we established that both the spectral position of the phase singularity in the fundamental STOV and the nonlinear phase-matching conditions influence the spectrum of the generated harmonic STOV fields. This provides precise tuning capability for UV STOV emissions. Our findings confirm a universal conservation law for spatiotemporal topological charge in nonlinear frequency conversion processes, offering a straightforward methodology for generating STOV fields across the entire visible and ultraviolet spectral ranges.

**Method**

In our experiments, we employed a 4f pulse shaper to generate fundamental-frequency STOV light fields. A commercial femtosecond laser system (Spectral-Physics SOL-35F-1K-HP-T) provided 35-fs pulses centered at 800 nm with a maximum pulse energy of 7 mJ at a repetition rate of 1 kHz. As show in Fig. 1, the pulses were first spatially dispersed by a grating (G1) at an incidence angle of 46°, then focused by a cylindrical lens (CL1; f = 150 mm) to perform Fourier transformation. At the focal plane of CL1, a spatial light modulator (SLM; Hamamatsu X13138) imprinted a spiral phase pattern in the momentum-spatial (*x*-*y*) plane with variable topological charge. The pulse reflected back from the SLM traversed CL1 and underwent inverse Fourier transformation before a second reflection on G1, exiting the pulse shaper via reflection from a beam splitter (BS). A convex lens (f = 1000 mm) subsequently performed Fourier transformation to generate the far-field STOV. For second-harmonic generation (SHG), we utilized type I BBO crystals (θ = 29.2°) with thicknesses of 0.3, 0.5, and 1 mm to convert linearly polarized STOV fields. For third-harmonic generation through sum-frequency generation (SFG), we cascaded a type I BBO crystal (θ = 43.0°) after the SHG stage, optimizing both crystals' rotation angles to control field polarizations and maximize third-harmonic conversion efficiency. We characterized the spatiotemporal topological charge using a diffraction-based method [23], directing the light field onto a diffraction grating (1200 lines/mm) and focusing with a cylindrical lens (CL2; f = 100 mm), then recording the diffraction patterns at the focal plane using a CCD camera. Optical filters isolated wavelength components during measurements, with distances L1 and L2 between cylindrical lens and grating/SLM maintained at f = 150 mm.

**Data availability**

All the data in this study are provided within the paper and its supplementary information. The data of this study are available from the corresponding author upon request.

**Acknowledgements**

This work was supported by the National Key Research and Development Program of China (2022YFE0107400), National Natural Science Foundation of China (11904232, 12034013, 62475158), Shanghai Science and Technology Commission (22ZR1444100). Authors are grateful to Prof. Qiwen Zhan, Dr. Liang Xu and Jiahao Dong of University of Shanghai for Science and Technology for helpful and enlightening discussions.

**Author contributions**

Y.L., Q.L. and G.S. proposed the original idea and initiated this project. Q.L., Q.Z., D.W., E.Z. and M.W. designed and performed the experiments. Q.L., Y.L., and G.S. analyzed the data. J.H. and J.W. completed the simulations. Q.L., Y.L. and G.S. prepared the manuscript. Q.L. and Y.L. supervised the project. All authors contributed to the discussion and writing of the manuscript.

**Competing interests**

The authors declare no competing interest.

**Figures**

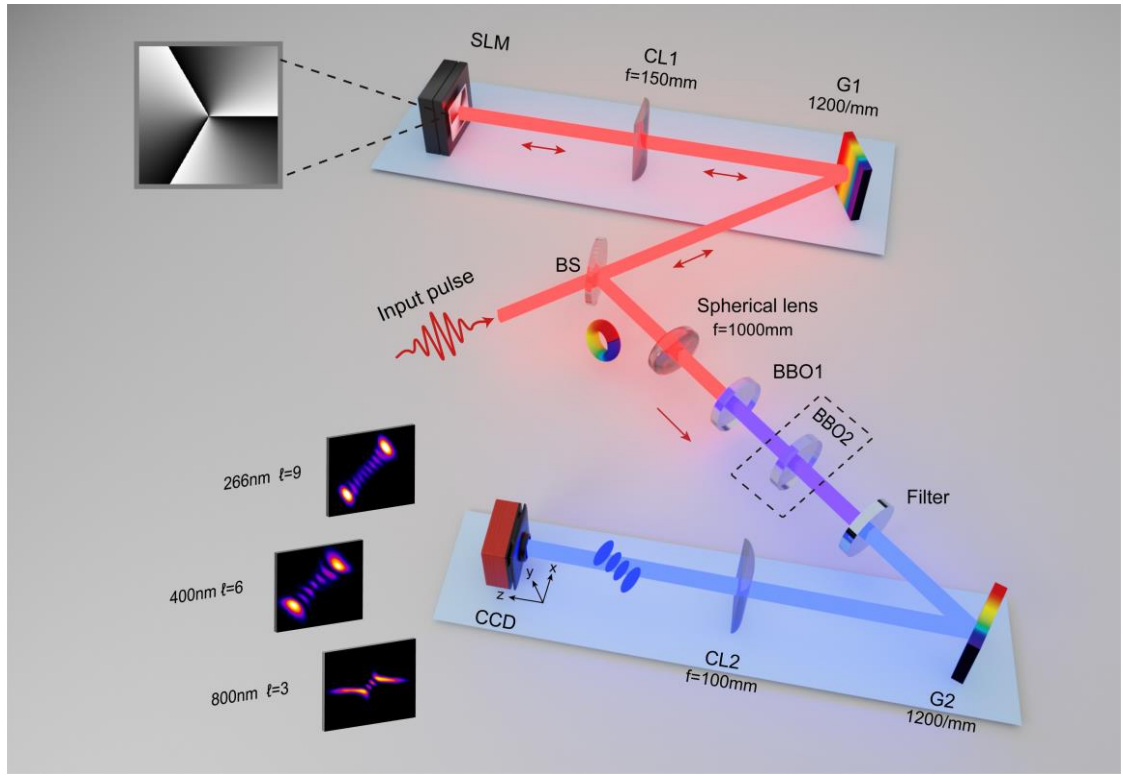

Fig. 1. **Experimental setup for generation of harmonics of the incident STOV field**. The 800 nm femtosecond laser pulses were diffracted by a grating (G1) and Fourier transformed by a cylindrical lens (CL 1), then shine on a spatial light modulator (SLM) where the phase singularity was implemented. The reflected pulse from the SLM transmits through the L1 and the G1 and exit from the pulse shaper by reflection from the beam splitter (BS), and a STOV field at 800 nm was formed with the Fourier transformation of the spherical lens. The second harmonic generation BBO crystal (BBO 1) and the sum frequency generation BBO crystal (BBO 2), as well as some spectral filters consititute the unit for frequency conversion of the STOV field. For characterization of the topologicial charge of the harmonic STOV field, another grating (G 2) and a cylindrical lens (CL 2) was employed, with the diffraction pattern recorded by a charge-coupled-device (CCD).

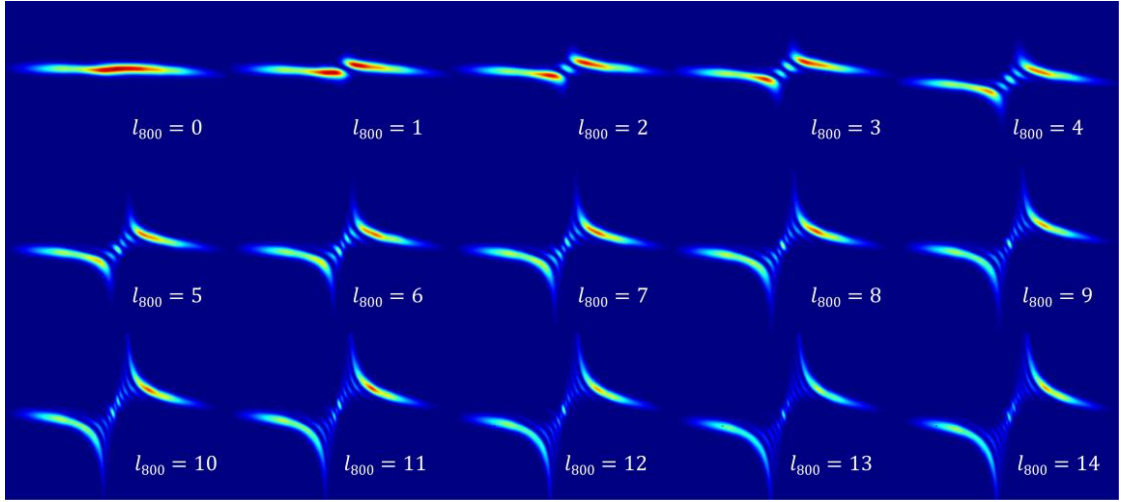

Fig. 2. **Experimental observation of STOV field at the 800 nm fundamental frequency**. The topological charge of the STOV fields are denoted in each panel.

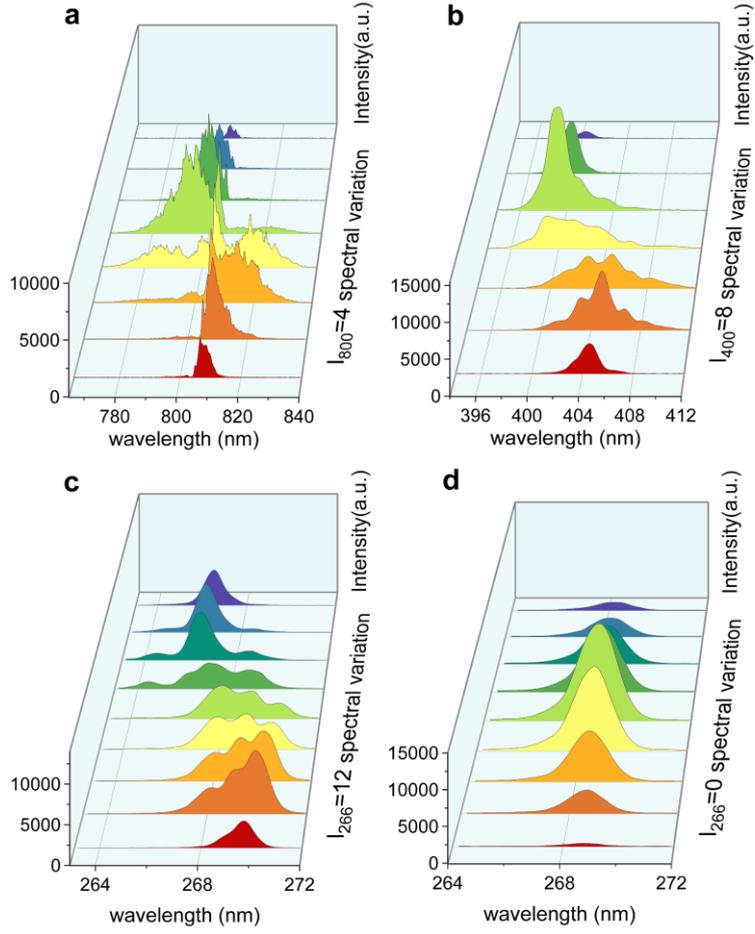

**Fig. 3. Spectrum measurement across the transverse direction of the STOV beam of different wavelength. a-c**, results for 800, 400, and 266nm, with the respective topological charge of 4, 8, 12. **d**, the spectrum measurement of 266 nm field with $l^{(3\omega)} = 0$.

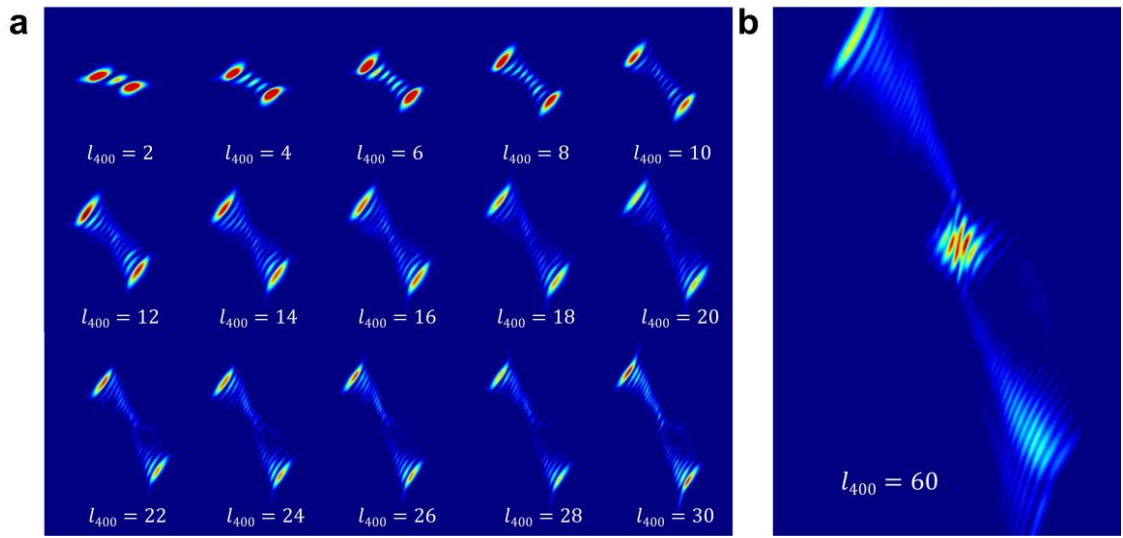

**Fig. 4. Second harmonic STOV field generation. a**, The experimental diffraction pattern of the generated second harmonic STOV at 400 nm with different topologic charge. In **b**, the corresponding results for $l^{(2\omega)} = 60$ is highlighted in enlarged view.

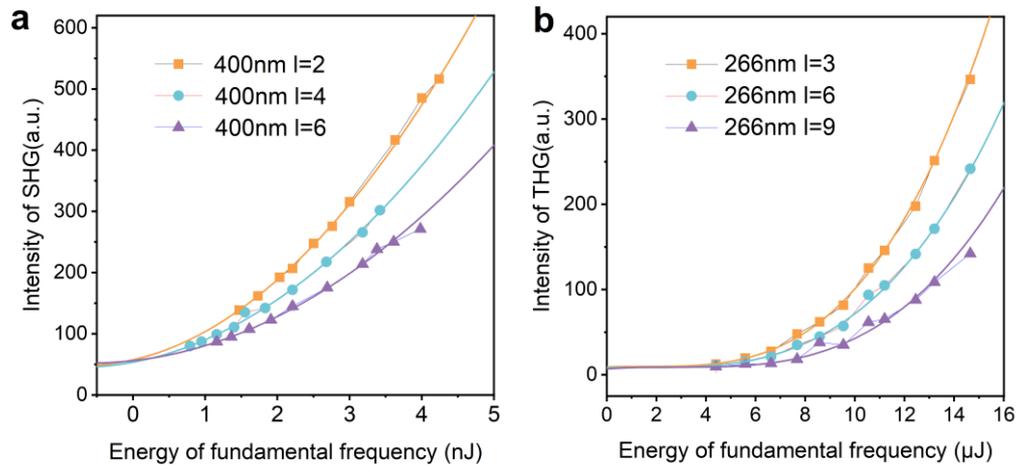

**Fig. 5. Energy scaling of the second and third harmonic STOV field. a**, The energy of the second harmonic STOV field with different topological charges with respect to the incident energy of the 800 nm STOV field. **b**, the results for the third harmonics STOV field at 266 nm. The solid curve in the two panels denote quadratic and cubic fitting.

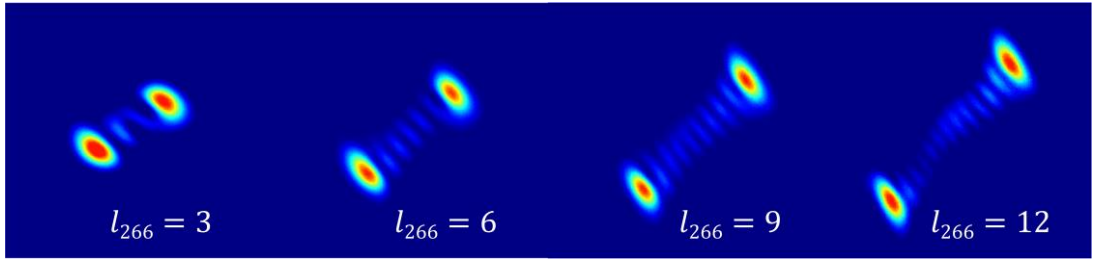

**Fig. 6. Cascaded generation of the third harmonic STOV field.** The experimental diffraction pattern of the 266 nm STOV field with topological charge $l^{(3\omega)} = 3, 6, 9, 12$.

# Supplementary information: Cascaded frequency conversion of highly charged femtosecond spatiotemporal optical vortices

Qingqing Liang,[1] Qiyuan Zhang,[1] Dan Wang,[1] Enliang Zhang,[1] Jianhua Hu,[1] Ming Wang,[1] Jinxin Wu,[1] Grover A. Swartzlander,[2] Yi Liu,[1, *]

[1]Shanghai Key Lab of Modern Optical System, University of Shanghai for Science and Technology, 516, Jungong Road, 200093 Shanghai, China

[2]Chester F. Carlson Center for Imaging Science, Rochester Institute of Technology, Rochester, New York, USA

* Corresponding author: yi.liu@usst.edu.cn.

Here we present comprehensive simulations and experimental results that further illustrate our capacity to precisely manipulate spatiotemporal optical vortices (STOVs) during nonlinear processes. These data highlight two pivotal advancements: (i) broadband wavelength tunability achieved by spectral singularity engineering, and (ii) increased conversion efficiency through crystallographic phase-matching optimization in BBO nonlinear crystals.

**Section 1: Theoretical Validation and Experimental Benchmarking of STOV Generation**

We confirmed our STOV generating approach through simulations and experiments. Employing the space-time coupling framework [1,2], we calculated diffraction patterns for STOV fields with topological charges from $l^{(\omega)}$= 0 to 14 (Supplementary Fig. 1). These simulations incorporated actual experimental parameters, including the 35-fs pulse duration and 4 mm beam width. Experimentally, our 4f pulse shaper using SLM produced well-defined 800 nm STOV fields. Their diffraction patterns consistently produced black fringes (Fig. 2), matching our calculations completely across all topological charges. Spectral measurements found important differences: conventional Gaussian pulses ($l^{(\omega)}$=0) maintained position-independent spectra centered at 800 nm. In contrast, STOV fields ($l^{(\omega)}$=4) exhibited distinct spatial-spectral chirp across the beam profile (Fig. 3a; Supplementary Fig. 2). This pattern coincides with theoretical expectations of spatiotemporal phase gradients [2], demonstrating both our technique's reliability and the intrinsic space-time coupling in STOV fields.

**Section 2: Nonlinear Crystal Thickness-Dependent Topological Charge Conservation**

In this section, we explored how BBO crystal thickness impacts topological charge preservation during harmonic conversion. For second-harmonic generation (SHG) with 0.3-mm-thick crystals, diffraction patterns preserved crisp features with excellent contrast ($\gamma$=0.92 ± 0.03 at $l^{(2\omega)}$=20), demonstrating good topological charge conservation. However, thicker 1.0-mm crystals dramatically reduced contrast ($\gamma$=0.61 ± 0.05). We explain this blurring (Supplementary Fig. 3) to spatiotemporal astigmatism produced by group velocity mismatch ($\Delta v_g = v_g^{(800)} - v_g^{(400)}$ =0.87 mm/ps), where temporal walk-off affects the harmonic field's phase structure. Third-harmonic generation demonstrated greater sensitivity to crystal thickness. With 1-mm crystals, the 299-fs temporal walk-off showed substantial pattern degradation for $l^{(3\omega)} \geq 15$ (Supplementary Fig. 4). Detailed research indicated these fuzzy patterns relate to mixed topological charges ($l^{(3\omega)} = 3l^{(\omega)} \pm \delta$) due to partial decoherence between fundamental and harmonic fields. These findings show crystals should meet $d_{max} < \tau_{FWHM}/(2\Delta v_g)$ to sustain topological integrity.

**Section 3: Spectral-Temporal Control of Harmonic STOV Wavelengths**

We tuned harmonic wavelengths by altering the spectral position of the phase singularity. Shifting the vortex center throughout the basic spectrum (786–819 nm in ~2.5 nm steps) while optimizing the BBO angle (~0.2°/nm offset) regulated phase matching. As seen in Supplementary Fig. 5, this asymmetrically transformed fundamental STOV diffraction patterns while keeping high-quality second harmonics. The harmonic wavelength shifted linearly with the singularity position ($\Delta\lambda^{(2\omega)} \approx 2\,\Delta\lambda^{(\omega)}$), covering 393–409.5 nm without impacting topological charge conservation ($l^{(2\omega)} = 2l^{(\omega)}$) or diffraction clarity. This tunability—achieved through coordinated spectral-phase control and nonlinear optics—enables wavelength-specific UV STOV production for applications like OAM communications and ultrafast spectroscopy.

**Figures**

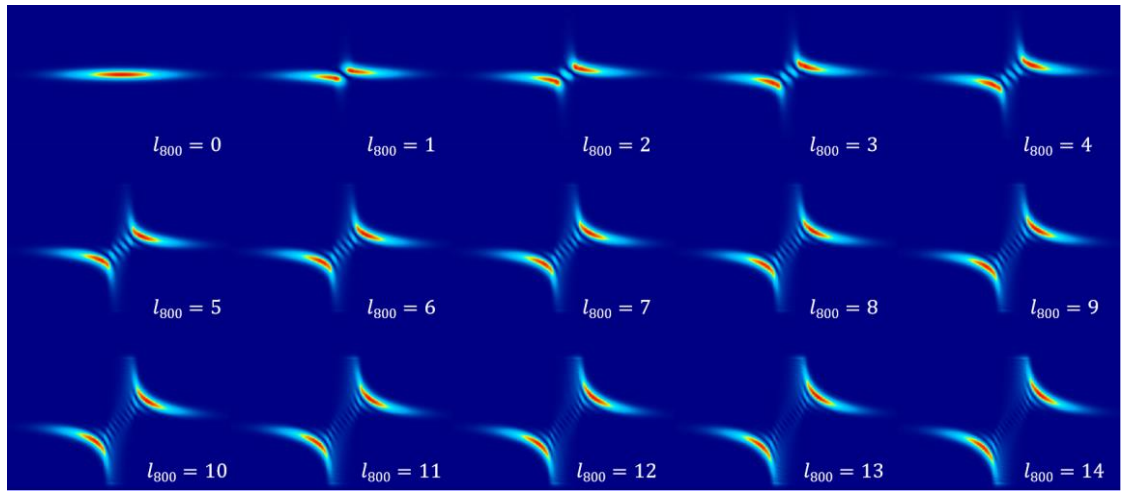

**Supplementary Fig. 1. Calculated diffraction patterns of the 800 nm STOV field with different topological charges.** With the calculated parameters chosen to those of the experiments.

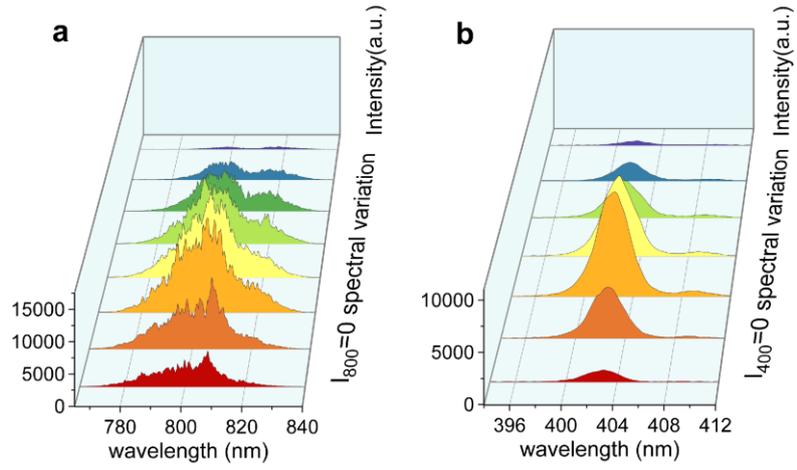

**Supplementary Fig. 2. Spectrum measurement across the transverse direction of the 800 nm and 400 nm beam without topological charge. a**, results for 800 nm field with $l^{(\omega)} = 0$. **b**, results for 400 nm field with $l^{(2\omega)} = 0$.

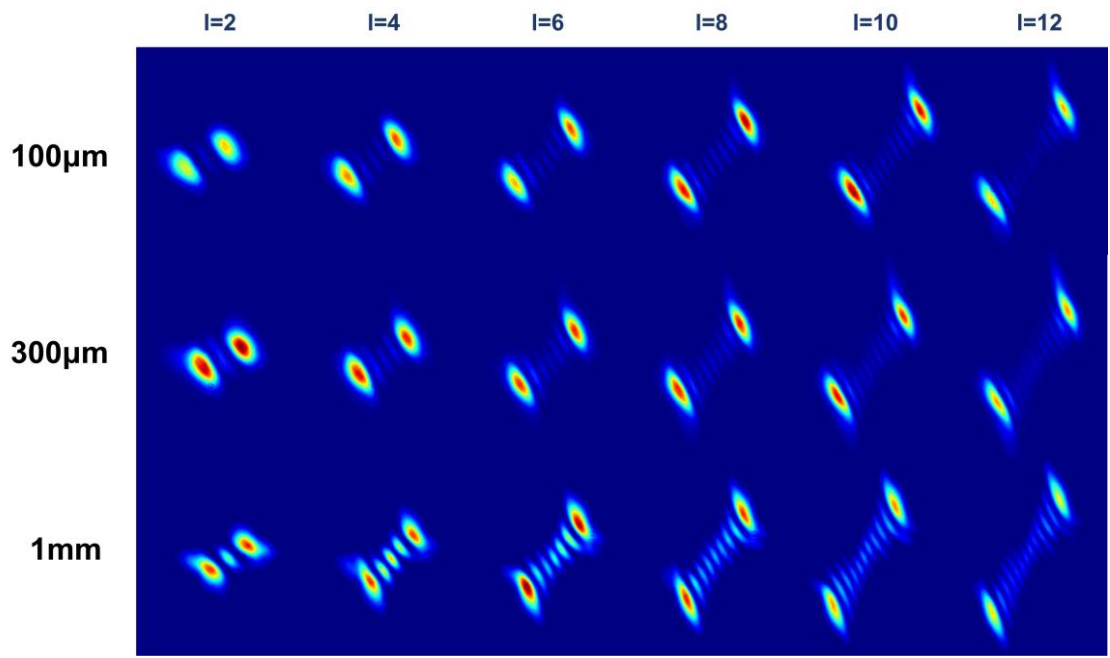

**Supplementary Fig. 3. Diffraction patterns of the STOV field for different thickness of SHG crystals.** The topological charge and the thickness of the crystals are denoted in the figure. With crystal of 1 mm, the diffraction patterns become blurred due to the spatiotemporal astigmatism.

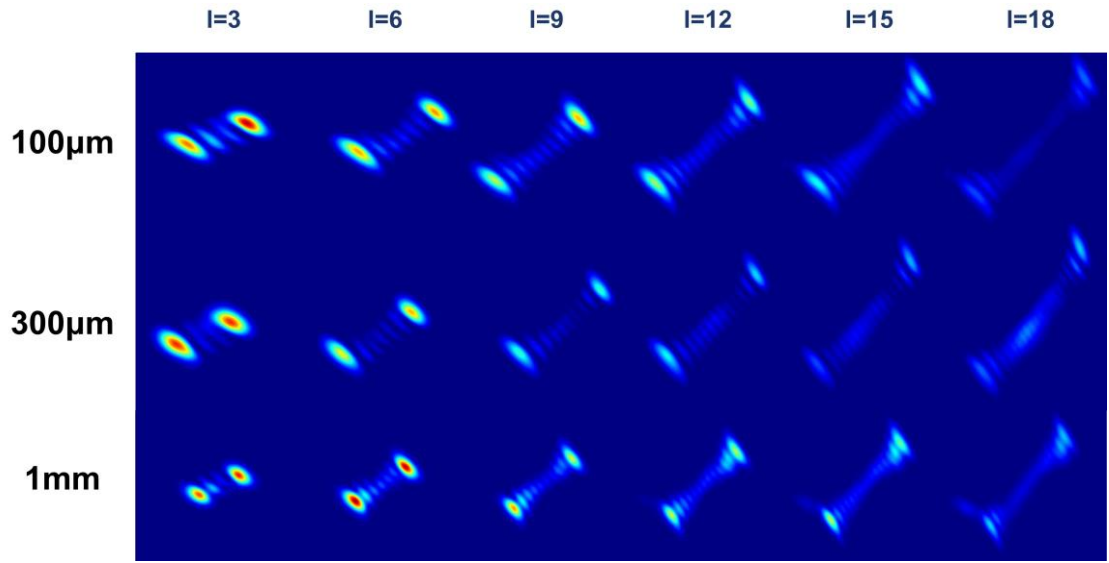

**Supplementary Fig. 4. Diffraction patterns of the third harmonic STOV field at 266 nm for different thickness of SFG crystals.** The topological charge and the thickness of the crystals are denoted. For higher topological charges and with SFG crystal of 1 mm, the diffraction patterns become blurred due to the spatiotemporal astigmatism and temporal walkoff inside the SFG crystal.

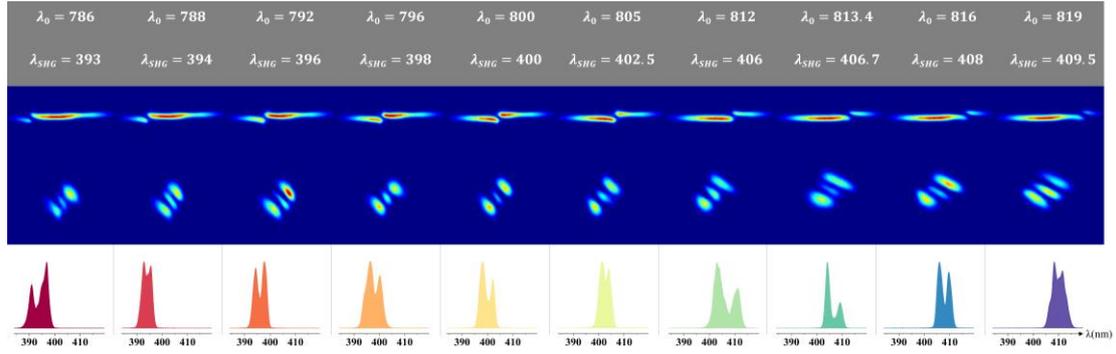

**Supplementary Fig. 5. Tunability of the second harmonic STOV field.** Top row: the diffraction patterns of the 800 nm STOV field with the topological charge located at varying wavelength, with the corresponding wavelength indicated in the tables above. Second row: the corresponding 400 nm STOV field. Third row : the measured spectrum of the second harmonic STOV field, with its central wavelength tuned from 393 nm to 409.5 nm. The numbers above the figure present the central wavelength where the phase singularity was applied on the SLM and the central wavelength of the second harmonic STOV.